\documentclass[Journal]{IEEEtran}
\usepackage{stmaryrd}
\usepackage{amsfonts}
\usepackage{amssymb}
\usepackage{cite}
\usepackage{graphicx}
\usepackage{psfrag}
\usepackage{subfigure}
\usepackage{amsmath}
\usepackage{array}
\begin{document}

\title{A Low-Overhead Energy Detection Based Cooperative Sensing Protocol for Cognitive Radio Systems}

\newtheorem{Thm}{Theorem}
\newtheorem{Lem}{Lemma}
\newtheorem{Cor}{Corollary}
\newtheorem{Def}{Definition}
\newtheorem{Exam}{Example}
\newtheorem{Alg}{Algorithm}
\newtheorem{Prob}{Problem}
\newtheorem{Rem}{Remark}

\author{\authorblockN{Shunqing Zhang, Tianyu Wu, and Vincent K. N. Lau\\}
\authorblockA{Department of Electronic and Computer Engineering\\
Hong Kong University of Science and Technology \\
Clear Water Bay, Kowloon, Hong Kong\\
Email: \{eezsq,wuty\}@ust.hk, eeknlau@ee.ust.hk}
\thanks{This work is supported by Research Grants Council (RGC) No. 615606.}
}

\maketitle


\begin{abstract}
Cognitive radio and dynamic spectrum access represent a new paradigm shift in more effective use of limited radio spectrum. One core component behind dynamic spectrum access is the sensing of primary user activity in the shared spectrum. Conventional distributed sensing and centralized decision
framework involving multiple sensor nodes is proposed to enhance the sensing performance. However,
it is difficult to apply the conventional schemes in reality since the overhead in sensing measurement and sensing reporting as well as in sensing report combining limit the number of sensor nodes that can participate in distributive sensing. In this paper, we shall propose a novel, low
overhead and low complexity energy detection based cooperative sensing framework for the cognitive
radio systems which addresses the above two issues. The energy detection based cooperative sensing
scheme greatly reduces the quiet period overhead (for sensing measurement) as well as sensing
reporting overhead of the secondary systems and the power scheduling algorithm dynamically allocate the transmission power of the cooperative sensor nodes based on the channel statistics of the links to the BS as well as the quality of the sensing measurement. In order to obtain design insights, we also derive the asymptotic sensing performance of the proposed cooperative sensing framework based
on the mobility model. We show that the false alarm and mis-detection performance of the proposed
cooperative sensing framework improve as we increase the number of cooperative sensor nodes.
\end{abstract}

\IEEEpeerreviewmaketitle
\section{Introduction}
\label{sect:intro}

Cognitive radio and dynamic spectrum access are important emerging technologies
\cite{Mitola99,Haykin05,Bhargava07} which may represent a new paradigm shift in more effective use
of limited radio spectrum. For instance, some of the license spectrum (such as UHF/VHF band) is
under-utilized \cite{Brodersen04,FCC04} and this motivated the standardization of Wireless Regional Area Network (WRAN) in IEEE 802.22 \cite{WRAN06} to exploit unused spectrum dynamically. One
important technical challenge in realizing the vision of cognitive radio systems is to maintain and control potential interference to primary users in the licensed spectrum \cite{Cadambe07,Devroye06}. There are in general two approaches in cognitive radio systems to
realize efficient spectrum sharing, namely a {\em static approach} and a {\em dynamic approach}. In the static approach (more conservative), the transmit power of the secondary system is limited such that the worst case interference to the primary users is controlled to an acceptable target
\cite{Hong08,Hamdi07} (protection contour) regardless of the instantaneous activity of the primary
system. While this approach is simple, it failed to exploit the dynamic activity and the location
of the primary users. In the dynamic approach, the secondary system will sense the activities of the primary system and transmits only during the activity gaps of the primary system. This approach allows better utilization of the shared spectrum especially when the primary system has bursty
traffic. As a result, this dynamic approach is commonly adopted as the basic framework in cognitive radio systems.

One core component behind dynamic spectrum access is the sensing of primary user activity in the
shared spectrum. For instance, the accuracy of sensing reports from secondary nodes are critical to the functioning of the cognitive radio systems. Imperfect sensing may cause either {\em false
alarm} or {\em mis-detection} problems. When we have false alarm in the sensing report, the
secondary system will be over-conservative because it falsely assumed the primary user is active.
When we have mis-detection in the sensing report, the secondary system may induce strong
interference to the primary users in the shared spectrum. In practice, providing an accurate
sensing measurement on the activity of primary users may be challenging for the following reasons.
For instance, the secondary system may not have knowledge of the signal structure as well as
channel states of the primary users. Hence, the sensing detection is usually done in a non-coherent manner \cite{Sahai04,Tandra08}. In addition, in order to maintain a reasonably low interference
level to primary system, the secondary nodes shall be able to detect the existence of very weak
primary signals \cite{Sahai04,Tandra08}. To improve the accuracy of the dynamic spectrum sensing,
classical detection theory suggests that the sensing performance can be improved by increasing the sensing time \cite{Tandra08,Kay98,Poor98}.  The sensing performance can be further enhanced by
distributive sensing involving multiple sensor nodes. In \cite{Unnikrishnan08,Visotsky05,Mishra06,Anandkumar07,Liu07}, a distributed sensing and centralized decision framework has been proposed. In \cite{Visotsky05}, each sensor node makes a hard-decision on the primary user activity and feedback the local decisions to the base station. The base station (BS) makes a final decision by majority voting. In \cite{Mishra06}, the authors proposed a similar
framework except that each sensor node feedback the soft sensing measurement and combined at the
BS. This framework is also adopted in the IEEE 802.22 sensing architecture.  In \cite{Tandra08}, an asymptotic relationship between the target non-coherent sensing performance and the number of
sensing samples needed was derived in the low SNR regime. In \cite{Anandkumar07,Liu07}, the authors proposed type-based distributed sensing schemes for the multi-access channels as well as wireless
sensor networks.

While distributive sensing is a promising solution to enhance the sensing performance in cognitive
radio systems, there are several important open issues to be addressed as elaborated below.
\begin{itemize}
\item {\bf Overhead in sensing measurement and sensing reporting}
In the above works, the overheads in sensing measurement and sensing reporting were ignored. For
instance, in 802.22 systems, the BS shall poll a consumer premise equipment (CPE) to perform
sensing measurement and report the sensing result in a round robin fashion. Yet, each sensing
measurement incur system overhead in the sense that the entire secondary system has to remain
silent during the sensing measurement\footnote{We notice that through simple adjustment, the
overhead in the sensing measurement can be reduced but the overhead in sensing reporting is still an important issue.}. In addition, reporting of sensing results reliably requires additional overheads. In fact, the issue of sensing report overhead is not too related to hard-decision or soft-decision. Previous literatures have assumed that the reporting links are perfect and the one-bit hard decision result can be delivered to the BS perfectly. However, in practical systems, to reliably deliver one-bit sensing information (even for hard decision) actually involves a lot of overheads, including reliable channel coding, packet preamble, CRC checking bits etc\footnote{For example, in IEEE 802.22 standard, the actual sensing reports are delivered using bulk measure report (BLM-REP) messages (which is one type of MAC-management message occupying a time-frequency burst). The BLM-REP message involves additional protection bits for reliable transmission.}. Moreover, if we allow all the ¡®primary detected¡¯ sensor nodes to report the sensing measurement simultaneously, the BS may not be able to separate them since the packet collision may happens. In other words, the sensing measurement report (BLM-REP) messages from various sensors have to be delivered to the BS sequentially. Typical ways to enforce this is for the BS to poll the CPEs to report sensing measurement as implemented in the 802.22.

\item {\bf Overhead in sensing report combining}
In all the above works, the BS has to combine a large number of distributive sensing measurements from secondary CPEs in the cognitive radio systems. This will incur storage issues\footnote{To store the complex-valued observations and the channel states from all the sensor nodes is quite challenging in the practical systems.} as well as the processing complexity which scales linearly with the number of sensing measurements.
\end{itemize}

Recent discoveries propose pioneering methods to reduce the potential overhead based on the
threshold-based cooperative sensing \cite{Sun07,Lunden07} or sequential detection
\cite{Chaudhari08,Taherpour07} and to improve the throughput of the cognitive transmission
\cite{Peh07,Liang08}, but the reporting link is assumed to be perfect which is in general difficult to implement in practical systems. In \cite{Chen04,Jiang05}, the authors considered the
imperfect reporting channels and the energy detection problem with AWGN reporting channel is
addressed in \cite{Quan08}. In this paper, we shall propose a low overhead and low complexity
energy detection based {\em cooperative sensing framework} for the cognitive radio systems which
addresses the above issues. The proposed framework consists of two parts, namely the {\em
cooperative sensing scheme} and the {\em power scheduling algorithm}. The cooperative sensing
scheme greatly reduces sensing reporting overhead of the secondary systems. Unlike conventional
distributive sensing schemes, the required sensing reporting overhead do not scale with the number
of cooperative sensor nodes. Furthermore, the BS does not need extra complexity in combining
multiple sensing reports from the cooperative sensor nodes. The power scheduling algorithm
dynamically allocates the transmission power of the cooperative sensor nodes based on the channel
statistics of the links to the BS as well as the {\em quality} of the sensing measurement. In order
to obtain design insights, we also derive the asymptotic sensing performance of the proposed
cooperative sensing framework based on the mobility model. We show that the false alarm and
mis-detection performance of the proposed cooperative sensing framework improve as we increase the
number of cooperative sensor nodes. In the low SNR region, to achieve a target false alarm and
mis-detection probability, the number of sensor nodes required scales in the order of $\mathcal
O\big(\Sigma_r(1 + \sqrt{1 + \Sigma_s})^2/\rho_r^2\rho_s^2\big)$ where $\rho_r,\Sigma_r$ represents
the mean and variance of the average received SNR of the ``sensor node''-``BS'' links and
$\rho_s,\Sigma_s$ represents the mean and variance of the average received SNR for the ``primary
user''-``sensor node'' links respectively. We also compare our proposed cooperative sensing
scheme with conventional distributive sensing schemes. Under similar system overhead, the proposed
scheme can have substantial performance enhancement and has important practical significance.

The rest of the paper is organized as follows. Section~\ref{sect:chan_mod} introduces the system
model. In Section~\ref{sect:dis_sen}, we outline the low-overhead cooperative sensing  scheme  for
cognitive radio systems and discuss the problem formulation of the power allocation in
Section~\ref{sect:probfor}. We propose our power allocation algorithm by solving the optimization
problem in Section~\ref{sect:pow_sch} and derive the asymptotic performance analysis in
Section~\ref{sect:asym_perf}. In Section~\ref{sect:experiments}, we give some numerical results and
discussions. Conclusions are given in Section~\ref{sect:conclusion}.

\section{System Model}
\label{sect:chan_mod} 
\begin{figure}
\centering
\includegraphics[width = 2.5in]{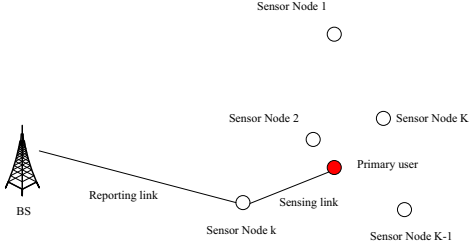}
\caption{System model of the proposed cooperative sensing framework
in the cognitive radio systems.} \label{fig:sys_mod}
\end{figure}

In this section, we shall elaborate on the channel model of the proposed
energy detection based cooperative sensing framework. Consider a cognitive radio system with a
single BS of secondary system and multiple sensor nodes as shown in Fig. \ref{fig:sys_mod}. Each
node is equipped with single antenna. To make it precise, we consider a cognitive radio system
containing $K$ sensor nodes, each making observations on a deterministic source signal $\theta$ of
the primary user within the licensed spectrum, where $\theta = 1$ when the primary user is active
and $\theta = 0$ otherwise. For notation convenience, we define the {\em $k^{th}$ sensing link} to
be the channel between the primary user and the $k^{th}$ sensor node and define the {\em $k^{th}$
reporting link} to be the channel between the $k^{th}$ sensor node and the BS as shown in Fig.
\ref{fig:sys_mod}.

Denote $h_{k}^{s}$ to be the channel coefficient of the $k^{th}$ sensing link. Each sensor node
observes for $T$ units of time and the time average received signal power can be described by
\begin{eqnarray}
|x_k|^2 = \frac{1}{T}\int_{t=0}^{T} {\big| h_{k}^{s} \theta x_p \sqrt{P_{pu}} + n_{k,t}^{s} \big
|}^{2} \textrm{d} t
\end{eqnarray}
where $h_{k}^{s}$ are Rayleigh fading coefficients with zero mean and variance $\Sigma_{k}^{s}$,
$x_p$ denotes the transmitted symbols of the primary user with normalized power and $P_{pu}$
denotes the transmitted power by the primary user. The noise $n_{k,t}^{s}$ is additive complex
Gaussian random variable with zero mean and normalized variance.

Denote $h_{k}^{r}$ to be the channel coefficient of the $k^{th}$ reporting link and $\mathcal X_k$
to be the transmitted signal at the $k^{th}$ sensor node. The corresponding received signal at the
BS is given by
\begin{eqnarray}
Y = h_{k}^{r}\mathcal X_k + n_r \label{eqn:sys_mod}
\end{eqnarray}
where $n_r$ is additive complex Gaussian random variable with zero mean and normalized variance,
and $h_{k}^{r}$ are Rayleigh fading coefficients with zero mean and variance $\Sigma_{k}^{r}$.

The following assumptions are made through the rest of the paper. Firstly, the $k^{th}$ sensor node
has perfect instantaneous channel state information (CSI) knowledge of the $k^{th}$ reporting link
through the transmitted preamble by the BS. Secondly, we assume the BS only has statistical
information (variances of the CSI) of the reporting links and the sensing links.

\section{Cooperative Sensing Scheme}
\label{sect:dis_sen} 
\begin{figure}
\centering
\includegraphics[width = 2.5in]{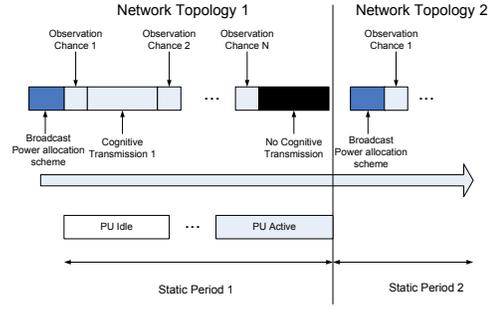}
\caption{A timing diagram comparison for the static period and the
observation chance.} \label{fig:time_rela}
\end{figure}

In this section, we shall describe the low overhead and low complexity energy
detection based cooperative sensing scheme. For easy illustration, we first define the {\em
observation chances} and the \emph{static period} as follows. An {\em observation chance} is
defined to be the time duration which allows all the sensor nodes to perform sensing and reporting
for one time. A \emph{static period} is defined to be the period when the channel statistics of the sensing links $\{\Sigma_{k}^{s}\}$ and the reporting links $\{\Sigma_{k}^{r}\}$ remain the same,
which contains many observation chances as shown in Fig. \ref{fig:time_rela}.

Consider a BS monitoring the behavior of the primary user with the help of $K$ sensor nodes. During each static period, the BS shall sense the primary user activity on the current channel. The
proposed cooperative sensing scheme can be briefly described as follows.

\begin{description}
\item [\emph{Step 1}]
In each static period, the BS shall determine the optimal power allocation scheme\footnote{The
rigorous definition of the power allocation scheme will be given in Section~\ref{sect:probfor}.}
and the corresponding threshold based on the channel statistics of the reporting links and the
sensing links. Before the BS of the secondary system transmits, it shall broadcast the downlink sensing request message (the power allocation scheme shall be only broadcast once for each static period and the preamble shall be delivered for each observation chance) to $K$ sensor nodes.
\item [\emph{Step 2}]
Upon receiving the request message, each sensor node shall sense the activity of the primary user for $T$ units of time and obtain a local sensing measurement of $|x_k|^2$ and the instantaneous CSI knowledge of the reporting link from the preamble.
\item [\emph{Step 3}]
Using the estimated CSI knowledge between the BS and the sensor nodes, all the $K$ sensor nodes amplify and forward a pre-equalized version of the sensing measurement to the BS. Specifically, the
transmitted signal at the $k^{th}$ sensor node, $\mathcal X_k$, is given by
\begin{eqnarray}
\mathcal X_k = \textrm{conj}(h_k^r) |x_k|^2 \sqrt{\alpha_k}
\end{eqnarray}
where $\textrm{conj}(\cdot)$ denotes the complex conjugate operation. The sensing results are then
RF combined in the air interface and the observation $X$ at the BS\footnote{In the network
initialization process, the BS groups the sensor nodes into the clusters and hence, knows about the
number of sensor nodes in each cluster. Since only those sensor nodes that are polled by the BS
shall upload measurement, the BS would know about $K$.} is given by
\begin{eqnarray}
X & = & \frac{\sum_{k=1}^{K}h_k^r \mathcal X_k + n_r}{K} \nonumber \\
& = & \frac{\sum_{k=1}^{K} |h_k^r|^2|x_k|^2 \sqrt{\alpha_k} + n_r}{K}
\end{eqnarray}
where $\alpha_k$\footnote{Precisely speaking, $\alpha_k$ shall be a function of the channel
statistics $\Sigma_{k}^{r}$ and $\Sigma_{k}^{s}$. For notation convenience, we use $\alpha_k$ to
represent $\alpha_k(\{\Sigma_{k}^{r}, \Sigma_{k}^{s}\})$ for all $k$ through the paper.} stands for
the amplify-and-forward (AF) gain of the $k^{th}$ sensor node.
\item [\emph{Step 4}]
Using the observation $X$, the BS shall determine the activity of the primary user. Since the BS
only has the statistical information of the reporting links and the sensing links, coherent
detection cannot be applied and we consider the \emph{envelop detection} (energy detection).
Specifically, the BS shall compare the observation result $X$ with a threshold $\mathcal T_0$ and
determine the activity of the primary user according to \eqref{eqn:comp_1}
\begin{eqnarray}
\hat{\theta} = g(X) = \left\{
\begin{array}{l l}
1, & X \geq {\mathcal T}_0 \\
0, & \textrm{otherwise}
\end{array}
\right. \label{eqn:comp_1}
\end{eqnarray}
\item [\emph{Step 5}]
The BS shall start the transmission in the current static period if the primary user is measured to be idle in the current observation chance and wait for the next observation chance
otherwise.
\end{description}

\begin{figure}
\centering
\includegraphics[width = 2.5in]{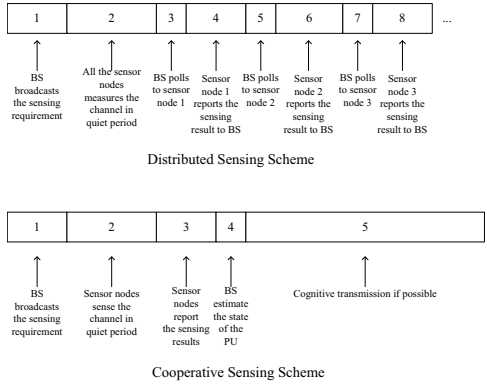}
\caption{Comparison between the traditional distributed sensing
scheme and the proposed cooperative sensing scheme.}
\label{fig:time_dia}
\end{figure}

In the current literatures, the conventional distributed sensing scheme requires the BS to poll each
of the sensor nodes and collect the corresponding sensing reports in a round robin fashion. The BS
shall then combine the sensing reports from different sensor nodes to make the final decision. Fig.
\ref{fig:time_dia} compares the timing diagram of the proposed cooperative sensing scheme and the
conventional distributed sensing scheme. Comparing with the conventional distributed sensing
scheme, the proposed cooperative sensing scheme has the following advantages.

\begin{itemize}
\item {\em Reduced System Overhead of Reporting.}
In the conventional distributed sensing scheme, the BS applies the polling scheme to get the
reports from multiple sensor nodes. Reporting overhead is proportional to the number of active
sensor nodes participating in the sensing. Using the proposed cooperative sensing scheme, the
number of active sensor nodes in the cooperative sensing system is no longer limited by the
reporting signaling overhead since the sensor nodes in the proposed scheme report the result
simultaneously. Hence, the proposed scheme can support much more active sensor nodes than the
conventional distributed sensing scheme with the same system overhead.
\item {\em Reduced System Overhead in Quiet Period.} Using the proposed cooperative sensing
scheme, independent measurements from multiple sensor nodes can be exploited. As a result, for the same performance as the conventional distributed sensing scheme, the system quiet period can be substantially reduced.
\item {\em Reduced Overhead in Sensing Report Combining.} To achieve the benefits of the
multiple sensor nodes, the distributed sensing scheme shall combine all the sensing reports from different sensor nodes using Maximum Ratio Combining (MRC). The processing complexity is proportional to the number of sensor nodes in the distributed sensing scheme. Using the proposed cooperative sensing scheme, the MRC is done automatically over the radio interface during the reporting phase of sensing measurement. As a result, no processing complexity at the BS is incurred.
\end{itemize}

\section{Problem Formulation}
\label{sect:probfor} Based on the proposed energy detection based cooperative sensing scheme, we
shall first define the system performance and then formulate the power scheduling problem as a
multi-object optimization problem.

\subsection{Definition of System Performance}
Since the power allocation is done at the BS side, it can only be a
function of the statistics of CSI information, i.e. the variances of
the reporting links $\{\Sigma_{k}^{r}\}$ and the variances of the
sensing links $\{\Sigma_{k}^{s}\}$. For notation convenience, we
first have the following definition on {\em power allocation (AF
gain) Scheme} and the system performance, namely, the {\em
probability of mis-detection} and the {\em probability of false
alarm} as follows.

\begin{Def}[Power Allocation (AF Gain) Scheme]
\label{Def:pow_all_po} A power allocation (AF gain) scheme $\mathbf
a$ is defined as the AF gain coefficients assigned for $K$ active
sensor nodes. Mathematically, the power allocation (AF gain) scheme
can be written as $\mathbf a = [\alpha_1, \alpha_2, \ldots,\alpha_K
]^{T}$. The power allocation (AF gain) vector $\mathbf a$ has to
satisfy the following transmission power constraint.
\begin{eqnarray}
&& \mathbb E_{\mathbf H} \big[|\mathcal X_k|^2 \big] = \mathbb
E_{\mathbf H} \big[|h_k^r|^2 |x_k|^4 \alpha_k \big] \leq P_k,\
\forall k
\end{eqnarray}
where $\mathbf H = \{h_k^r,h_k^s\}$ denotes the instantaneous CSI
information of the reporting links and the sensing links.
\end{Def}

\begin{Def} [Probability of Mis-Detection]
\label{Def:ave_perf} Given the statistics of CSI information
$\{\Sigma_{k}^{r}, \Sigma_{k}^{s}\}$, the power allocation (AF gain)
scheme $\mathbf a$ and the threshold $\mathcal T_0$, the
\emph{probability of mis-detection} is defined to be the probability
that the BS cannot detect the transmission of primary user when the
primary user is in active state. Mathematically, the probability of
mis-detection $P_{MD}$, as a function of power allocation (AF gain)
scheme $\mathbf a$ and the threshold $\mathcal T_0$, can be
described as
\begin{equation}
P_{MD} (\mathbf a,\mathcal T_0) = Pr \big(\hat{\theta} = 0 | \theta
= 1, \{\Sigma_{k}^{r},\Sigma_{k}^{s}\}, \mathbf a, \mathcal T_0
\big)
\end{equation}
\end{Def}

\begin{Def} [Probability of False Alarm]
\label{Def:ave_perf} Given the statistics of CSI information
$\{\Sigma_{k}^{r}, \Sigma_{k}^{s}\}$, the power allocation (AF gain)
scheme $\mathbf a$ and the threshold $\mathcal T_0$, the
\emph{probability of false alarm} is defined to be the probability
that the BS reports the transmission of primary user erroneously
when the primary user is in idle state. Mathematically, the
probability of false alarm $P_{FA}$, as a function of power
allocation (AF gain) scheme $\mathbf a$ and the threshold $\mathcal
T_0$, can be described as
\begin{equation}
P_{FA}(\mathbf a, \mathcal T_0) = Pr \big(\hat{\theta} = 1 | \theta
= 0, \{\Sigma_{k}^{r}, \Sigma_{k}^{s}\}, \mathbf a, \mathcal T_0
\big)
\end{equation}
\end{Def}

The events of mis-detection and false alarm are undesired in the
cognitive radio systems, since the mis-detection causes the
interference to the primary user which is not allowed and false
alarm reduces the opportunities of utilizing the unoccupied
channels. Hence, how to reduce the average probability of
mis-detection and false alarm simultaneously is the major objective
of this paper. In the next section, we shall formulate the power
allocation design into an multi-objective optimization problem.

\subsection{Problem Formulation}
In order to minimize the average probability of mis-detection and
false alarm simultaneously, we shall first calculate the probability
of mis-detection and false alarm, which is summarized in the
following lemma.
\begin{Lem}
\label{lem:pe} For sufficiently large $K$\footnote{The theoretical results in Lemma 1 are basically an asymptotic result based on Central Limit Theorem. However, we see that in practice, the result
is quite accurate even for moderate $K \sim 20$.} and $T$, the probability of mis-detection and false alarm
is given by:
\begin{eqnarray}
Pr \big(\hat{\theta} = 0 | \theta = 1, \{\Sigma_{k}^{r},
\Sigma_{k}^{s}\}, \mathbf a, \mathcal T_0 \big)
= Q \big(\frac{\mu_1 - \mathcal T_0}{\sigma_1}\big) \\
Pr \big(\hat{\theta} = 1 | \theta = 0, \{\Sigma_{k}^{r},
\Sigma_{k}^{s}\}, \mathbf a, \mathcal T_0 \big) = Q \big( \frac{
\mathcal T_0 - \mu_0}{\sigma_0}\big)
\end{eqnarray}
with
\begin{eqnarray*}
\mu_0 & = & \frac{\sum_{k=1}^{K}\textrm{SNR}_{k}^{r}}{K} \\
\sigma_0^2 & = & \frac{3 \sum_{k=1}^{K} {\textrm{SNR}_{k}^{r,2}} + 1}{K^2} \\
\mu_1 & = & \frac{\sum_{k=1}^{K}\textrm{SNR}_{k}^{r}(\textrm{SNR}_{k}^{s} + 1)}{K} \\
\sigma_1^2 & = & \frac{3  \sum_{k=1}^{K} {\textrm{SNR}_{k}^{r}}^2 (
\textrm{SNR}_{k}^{s,2} + \frac{2}{3} \textrm{SNR}_{k}^{s} + 1)  +
1}{K^2}\nonumber
\end{eqnarray*}
where $\textrm{SNR}_{k}^{s} = P_{pu} \Sigma_k^s$ denotes the
received SNR at the $k^{th}$ node from the primary user,
$\textrm{SNR}_{k}^{r} = \sqrt{\alpha_k} \Sigma_k^r$ denotes the
ratio of the received signal over the transmitted signal from
$k^{th}$ sensor node, and $Q(\cdot)$ denotes the standard Gaussian
complementary c.d.f. \cite{Proakis00}.
\end{Lem}
\proof Please refer to Appendix \ref{pf:lem_pe} for the proof.
\endproof

Based on the relationship between the system performance (i.e. the
probability of missing and false-alarm), the power allocation (AF
gain) scheme and the threshold, the optimal power allocation (AF
gain) scheme $\mathbf a^{*}$ and the optimal threshold $\mathcal
T_0^{*}$ can be described by the following multi-objective
optimization problem\footnote{In this paper, we focus on study the
Pareto optimality as specified in the next section.}.
\begin{eqnarray}
(\mathbf a^{*}, \mathcal T_0^*) & = & \arg \min_{(\mathbf a, \mathcal
T_0)} \left(
\begin{array}{l}
P_{MD}\big(\mathbf a, \mathcal T_0\big) \\
P_{FA}\big(\mathbf a, \mathcal T_0\big)
\end{array}
\right) \nonumber \\
& = & \arg \min_{(\mathbf a, \mathcal T_0)} \left(
\begin{array}{l}
Q \big( \frac{\mu_1 - \mathcal T_0}
{\sigma_1}\big) \\
Q \big( \frac{ \mathcal T_0 - \mu_0}{\sigma_0}\big)
\end{array}
\right) \label{eqn:tar_1}
\end{eqnarray}

\section{Power Scheduling Algorithm}
\label{sect:pow_sch} In this section, our target is to find the optimal power allocation scheme
based on solving the optimization problem given by \eqref{eqn:tar_1}. To measure the efficiency of
the multi-objective situation, \emph{Pareto Optimality}, is widely used to describe those tradeoff
relations. A solution can be considered Pareto optimal if there is no other solution that performs
at least as good on every criteria and strictly better on at least one criteria\footnote{Other
restrictions on the probability of mis-detection or false alarm (e.g. $P_{MD}, P_{FA} \leq 0.1$)
can be addressed by a simple intersection operation between the tradeoff curve and the region
quantified by the constraints.}. Mathematically, $(\mathbf a^*, \mathcal T_0^*)$ is Pareto optimal
if we cannot find a solution $(\hat{\mathbf a}, \hat{\mathcal T_0)}$ such that
\begin{eqnarray}
\left(
\begin{array}{l}
P_{MD}\big(\hat{\mathbf a}, \hat{\mathcal T_0}\big) \\
P_{FA}\big(\hat{\mathbf a}, \hat{\mathcal T_0}\big)
\end{array}
\right) \preceq \left(
\begin{array}{l}
P_{MD}\big(\mathbf a^*, \mathcal T_0^*\big) \\
P_{FA}\big(\mathbf a^*, \mathcal T_0^*\big)
\end{array}
\right)
\end{eqnarray}
where $\preceq$ denotes the element-wise inequality. As a result, we
solve the multi-objective optimization problem \eqref{eqn:tar_1} by
characterizing the optimal trade-off curve, which is a set of Pareto
optimal values for a multi-criteria problem. To solve for Pareto
optimal solution, we can scalarize the multi-objective optimization
problem \eqref{eqn:tar_1} as follows \cite{Boyd03}
\begin{eqnarray}
& \min_{\mathbf a, \mathcal T_0 } & Q \big( \frac{\mu_1 - \mathcal
T_0}{\sigma_1}\big) + \beta Q \big( \frac{\mathcal T_0 -
\mu_0}{\sigma_0}\big)
\nonumber \\
& \textrm{s.t.} &  \mathbb E_{\mathbf H} \big[|h_k^r|^2 |x_k|^4
\alpha_k \big] \leq
P_k, \ \forall k , \quad \mathbf a \succeq 0 \label{eqn:tar_2}
\end{eqnarray}
where $\beta$ is the corresponding weight\footnote{The probability
of mis-detection describes how well the primary system can be
protected and the probability of false alarm mainly constraints the
utilization of cognitive transmission. The objective here can be
interpreted as a joint consideration of the above two criteria.},
which balances the tradeoff between the probability of mis-detection
and false alarm\footnote{Any solution to the problem
\eqref{eqn:tar_2} is Pareto optimal but not conversely.}.

\subsection{Optimal Value of the Threshold $\mathcal T_0^{*}$}
We shall first optimize the threshold $\mathcal T_0$ in problem
\eqref{eqn:tar_2}. Specifically, the optimal threshold $\mathcal
T_0^{*}$ is given by the following optimization problem.
\begin{eqnarray}
\mathcal T_0^{*} = \arg \min_{\mathcal T_0} \ Q \big(\frac{\mu_1 -
\mathcal T_0}{\sigma_1}\big) + \beta Q\big( \frac{\mathcal T_0 -
\mu_0}{\sigma_0}\big)
\end{eqnarray}

Since the above optimization problem is an unconstrained minimization problem, we can differentiate
the objective function with respect to (w.r.t.) the threshold $\mathcal T_0$ and calculate the
optimal value $\mathcal T_0^*$ by setting the first-order derivative equal to zero. As a
result, the optimal choice of the threshold $\mathcal T_0^{*}$ shall satisfy the following
relation. $\frac{1}{\sqrt{2 \pi}} \bigg( \exp\big(-\frac{1}{2}(\frac{\mu_1 -
\mathcal T_0^{*}}{\sigma_1})^2\big)( - \frac{1}{\sigma_1}) + \beta
\exp\big(-\frac{1}{2}(\frac{\mathcal T_0^{*} -
\mu_0}{\sigma_0})^2\big)\frac{1}{\sigma_0} \bigg) = 0$.
Equivalently, the optimal value of the threshold $\mathcal T_0^{*}$
has the following relations.
\begin{eqnarray}
\exp\big(-\frac{1}{2}(\frac{\mu_1 - \mathcal
T_0^{*}}{\sigma_1})^2\big) \frac{1}{\sigma_1} = \beta
\exp\big(-\frac{1}{2}(\frac{\mathcal T_0^{*} -
\mu_0}{\sigma_0})^2\big)\frac{1}{\sigma_0}
\end{eqnarray}
With some mathematical manipulation, we have
\begin{eqnarray}
(\frac{\mathcal T_0^{*} - \mu_0}{\sigma_0})^2 - (\frac{\mu_1 -
\mathcal T_0^{*}}{\sigma_1})^2 = 2 \ln(\frac{\beta
\sigma_1}{\sigma_0}) \label{eqn:thr_2}
\end{eqnarray}
The optimal threshold $\mathcal T_0^{*}$ is hence given by
\begin{eqnarray}
\mathcal T_0^{*} & = & \frac{\sigma_1^2 \mu_0 - \sigma_0^2 \mu_1}{\sigma_1^2 - \sigma_0^2} + \frac{\sigma_0 \sigma_1}{\sigma_1^2 - \sigma_0^2} \nonumber \\
& & \times \sqrt{(\mu_1 - \mu_0)^2 + 2(\sigma_1^2 -
\sigma_0^2) \ln(\beta \sigma_1/\sigma_0)}
\label{eqn:thre_1}
\end{eqnarray}
Substitute \eqref{eqn:thr_2} into the optimization problem
\eqref{eqn:tar_2}, we have
\begin{eqnarray}
& \min_{\mathbf a} & Q \Big(\sqrt{(\frac{\mathcal T_0^{*} -
\mu_0}{\sigma_0})^2 - 2 \ln(\frac{\beta \sigma_1}{\sigma_0})}\Big)
+ \beta Q\big( \frac{\mathcal T_0^{*} - \mu_0}{\sigma_0}\big) \nonumber \\
& \textrm{s.t.} &  \mathbb E_{\mathbf H} \big[|h_k^r|^2 |x_k|^4
\alpha_k \big] \leq
P_k, \ \forall k, \quad \mathbf a \succeq \mathbf 0, \nonumber \\
&& \textrm{Eqn.} \ \eqref{eqn:thre_1} \label{eqn:thr_3}
\end{eqnarray}

\subsection{Optimal Power Allocation Scheme}
The objective function in the optimization problem \eqref{eqn:thr_3} is a monotonic decreasing
function with respect w.r.t. $(\mathcal T_0^{*} - \mu_0)/\sigma_0$ for $\mathcal T_0^{*} \geq
\mu_0$. Hence, we can transform the original minimization problem into the following maximization
problem.
\begin{eqnarray}
& \max_{\mathbf a} & \frac{\mathcal
T_0^{*} - \mu_0}{\sigma_0} \nonumber \\
& \textrm{s.t.} &  \mathbb E_{\mathbf H} \big[|h_k^r|^2 |x_k|^4
\alpha_k \big] \leq P_k, \ \forall k , \quad \mathbf a \succeq \mathbf 0, \nonumber \\
&& \textrm{Eqn.} \ \eqref{eqn:thre_1} \label{eqn:thr_4}
\end{eqnarray}

In fact, to solve the optimization problem \eqref{eqn:thr_4} is non-trivial since the problem is
non-convex in $\mathbf a$. Consider the Lagrangian dual of \eqref{eqn:thr_4} given by \cite{Yu06}:
\begin{eqnarray}
L(\mathbf a, \mathbf \Gamma) =  \frac{\mathcal T_0^{*} - \mu_0}{\sigma_0} - \sum_{k=1}^{K}\gamma_k
\Big(\mathbb E_{\mathbf H} \big[|h_k^r|^2 |x_k|^4 \alpha_k \big] - P_k \Big) \nonumber
\end{eqnarray}
where $\mathbf \Gamma = [\gamma_1, \gamma_2, \ldots, \gamma_K]$ are the dual variables. Define the
dual objective $g( \mathbf \Gamma)$ as a maximization of the Lagrangian $ g(\mathbf \Gamma) =
\max_{\mathbf a} L(\mathbf a, \mathbf \Gamma)$. The dual optimization problem is
\begin{eqnarray}
\min_{\mathbf \Gamma} && g( \mathbf \Gamma) \nonumber \\
\textrm{s.t.} && \mathbf \Gamma \succeq \mathbf 0 \label{eqn:tar_3}
\label{eqn:dual_1}
\end{eqnarray}
While the standard way of solving constrained optimization problem
is to form a Lagrangian dual, it is important to make sure that the
duality gap between the original problem and the dual problem is
zero. We first have the following important theorem on the duality
gap of the problem.

\begin{Thm}
\label{Thm:pri_dual}The non-convex optimization problem
\eqref{eqn:thr_4} and its dual problem \eqref{eqn:tar_3} has a zero
duality gap \cite{Boyd03}, i.e., the primal problem
\eqref{eqn:thr_4} and the dual problem \eqref{eqn:tar_3} have the
same optimal value.
\end{Thm}
\proof Please refer to Appendix \ref{pf:thm_pri_dual} for the proof.
\endproof

Theorem \ref{Thm:pri_dual} establishes the relationship between the primal optimization problem
\eqref{eqn:thr_4} and its dual problem. Since it has a zero duality gap, we can solve the primal
optimization problem \eqref{eqn:thr_4} by solving its dual problem \eqref{eqn:tar_3}, which is a
standard convex optimization problem w.r.t. the Lagrangian variables $\mathbf \Gamma$. By
substituting the expression of $\mathcal T_0^*$, $g(\mathbf \Gamma)$ in the optimization problem
\eqref{eqn:tar_3} becomes
\begin{eqnarray}
& \max_{\mathbf a} & \frac{\sigma_0 (\mu_0 - \mu_1)}{\sigma_1^2 - \sigma_0^2} - \sum_{k=1}^{K}\gamma_k \Big(\mathbb E_{\mathbf H} \big[|h_k^r|^2 |x_k|^4 \alpha_k \big] - P_k \Big)\nonumber \\
&& + \frac{\sigma_1
\sqrt{(\mu_1 - \mu_0)^2 + 2(\sigma_1^2 - \sigma_0^2) \ln(\beta
\sigma_1/\sigma_0)}}{\sigma_1^2 - \sigma_0^2} \nonumber \\
& \textrm{s.t.} & \mathbf a \succeq \mathbf 0 \label{eqn:thr_5}
\end{eqnarray}

Introduce the slack variable $G$, such that $ \big(\sigma_0 (\mu_0 - \mu_1) + \sigma_1
\sqrt{(\mu_1 - \mu_0)^2 + 2(\sigma_1^2 - \sigma_0^2) \ln(\beta \sigma_1/\sigma_0)}\big)/(\sigma_1^2
- \sigma_0^2) \geq G$ and the optimization problem \eqref{eqn:thr_5} can be transferred as follows.
\begin{eqnarray}
& \max_{\mathbf a, G} & G - \sum_{k=1}^{K}\gamma_k \Big(\mathbb E_{\mathbf H} \big[|h_k^r|^2
|x_k|^4 \alpha_k \big] - P_k \Big) \\
& \textrm{s.t.} & \sigma_1^2 \big((\mu_1 - \mu_0)^2 + 2(\sigma_1^2 - \sigma_0^2)\ln(\beta
\sigma_1/\sigma_0)\big) \nonumber \\
&& \geq \big[G(\sigma_1^2
- \sigma_0^2) + \sigma_0(\mu_1 - \mu_0)\big]^2 \label{eqn:thr_7} \\
& & \mu_0 = \frac{\sum_{k=1}^{K} \textrm{SNR}_{k}^{r}}{K}, \sigma_0^2 = \frac{3\sum_{k=1}^{K} {\textrm{SNR}_{k}^{r}}^2 + 1}{K^2} \nonumber\\
& & \mu_1 = \frac{ \sum_{k=1}^{K} \textrm{SNR}_{k}^{r} (\textrm{SNR}_{k}^{s} + 1)}{K} , \nonumber \\
& & \sigma_1^2
= \frac{\sum_{k=1}^{K} {\textrm{SNR}_{k}^{r}
}^2(3\textrm{SNR}_{k}^{s,2} + 2\textrm{SNR}_{k}^{s} + 3) + 1}{K^2} \nonumber \\
& & \mathbf a \succeq \mathbf 0 \label{eqn:thr_6}
\end{eqnarray}

Due to the constraint \eqref{eqn:thr_7}, the above optimization problem is still non-convex. To simplify the problem, we can bound the original non-convex constraint  \eqref{eqn:thr_7} by a linear constraint w.r.t. $\mathbf a$, e.g.
$ (\frac{\mu_1 - \mu_0}{\sigma_0} - G)^2 \sigma_0^2 + \sigma_1^2\Big(G^2 - 2 \ln \beta + \ln (\frac{\sigma_1^2}{\sigma_0^2})\Big) \geq \mathbf C^{T}\mathbf a + d \geq 0 $,
where $\mathbf C$ and $d$ can be determined by $G$, $\beta$ and $\{\textrm{SNR}_k^s\}$ since $\sigma_0^2$ and $\sigma_1^2$ are linear in $\mathbf a$ and $\Big(G^2 - 2 \ln \beta + \ln (\frac{\sigma_1^2}{\sigma_0^2})\Big)$, $(\frac{\mu_1 - \mu_0}{\sigma_0} - G)^2 $ can be bounded for fixed $G$, $\beta$ and $\{\textrm{SNR}_k^s\}$. Moreover, since the objective function of the problem \eqref{eqn:thr_6} is non-increasing in $\mathbf a$\footnote{In this paper, the objective function $f(\mathbf a)$ is non-increasing in $\mathbf a$ means the gradient of the objective w.r.t. $\mathbf a$ is element-wise less than or equal to zero, i.e. $\nabla_{\mathbf a} f(\mathbf a) = [\frac{\partial f}{\partial \alpha_1} \frac{\partial f}{\partial \alpha_2} \cdots \frac{\partial f}{\partial \alpha_K}]^H \preceq \mathbf 0$.}, the original optimization problem \eqref{eqn:thr_6} can be well approximated\footnote{Notice that the solution of the approximated optimization problem is also an approximated solution of the original optimization problem \eqref{eqn:thr_6} as well.} by a standard quasi-convex optimization problem \cite{Boyd03} through the relaxization of the non-convex constraint \eqref{eqn:thr_7}. As a result, efficient algorithm such as bisection search algorithms \cite{Boyd03} can be used to solve this problem. By solving problem \eqref{eqn:thr_5}, we can obtain the value of $g(\mathbf \Gamma)$ and find the gradient of $g(\mathbf \Gamma)$ w.r.t. $\mathbf \Gamma$. Hence, problem \eqref{eqn:dual_1} can be solved through standard gradient search \cite{Boyd03} and finally we can obtain the optimal value of the
original problem \eqref{eqn:tar_2}.

\section{Asymptotic Performance Analysis}
\label{sect:asym_perf} In this section, we shall derive the probability of false alarm and
mis-detection averaged over multiple static periods based on a mobility model\footnote{In this
paper, we mainly focus on analyzing the probability of false alarm and mis-detection. The
performance advantages of the proposed cooperative sensing scheme can be translated into the
reduction of quite period under the same probability of false alarm and mis-detection as well.}. We
assume the channel statistics $\{\Sigma_k^r, \Sigma_k^s\}$ follows the statistic model\footnote{The
statistical fluctuations of $\{\Sigma_k^r, \Sigma_k^s\}$ is driven by the mobility of the CPE in
the secondary system.} with mean and variance of the received $\textrm{SNR}_{k}^{r}$ and
$\textrm{SNR}_{k}^{s}$ given by
\begin{eqnarray}
\mu_{\textrm{SNR}_{k}^{r}} & = & \mathbb E_{\Sigma_k^r}\big[\textrm{SNR}_{k}^{r}\big] = \sqrt{\alpha_k}
\Lambda_r, \\
\sigma^{2}_{\textrm{SNR}_{k}^{r}} & = & \mathbb E_{\Sigma_k^r}
\big[(\textrm{SNR}_{k}^{r} - \mu_{\textrm{SNR}_{k}^{r}})^2\big] = \Sigma_r, \label{eqn:mod_1}\\
\mu_{\textrm{SNR}_{k}^{s}} & = & \mathbb E_{\Sigma_k^s}\big[\textrm{SNR}_{k}^{s}\big] = P_{pu}
\Lambda_s, \\
\sigma^{2}_{\textrm{SNR}_{k}^{s}} & = & \mathbb
E_{{\Sigma_k^s}^2}\big[(\textrm{SNR}_{k}^{s} - \mu_{\textrm{SNR}_{k}^{s}})^2\big] = \Sigma_s, \label{eqn:mod_4}
\end{eqnarray}

Denote $P_e^{*}\big(\{\Sigma_k^r,\Sigma_k^s\}\big) = P_{MD}(\mathbf
a,\mathcal T_0) + \beta P_{FA}(\mathbf a, \mathcal T_0)$ to be the
probability of false alarm and mis-detection error performance of
the proposed cooperative sensing scheme under the channel statistics
$\{\sigma_k^r,\sigma_k^s\}$ and the optimal power allocation scheme
$\mathbf a^*$. Hence, the average probability of false alarm and
mis-detection error performance $\bar{P}_e^*$ is given by
\begin{eqnarray}
\bar{P}_e^* = \mathbb E_{\{\Sigma_k^r,\Sigma_k^s\}} \Big[
P_e^{*}\big(\{\Sigma_k^r,\Sigma_k^s\}\big) \Big]
\end{eqnarray}

We assume the power constraints are uniform over different sensor nodes, i.e. $P_k = P$ for all
$k$. Meanwhile, we consider an upper bound on the error probability of false alarm and
mis-detection by considering a constant AF gain policy, i.e., $\alpha_k = \alpha$ with $\alpha =
\frac{P}{\mathbb E_{\mathbf H} [|h_k^r|^2 |x_k|^4 ]}$. Since the constant AF gain is one of the
many schemes in the optimization domain, the error performance obtained is an achievable upper
bound $P_e^{u}\big(\{\Sigma_k^r,\Sigma_k^s\}\big)$ and given by
\begin{eqnarray}
& & P_e^{*}\big(\{\Sigma_k^r,\Sigma_k^s\}\big) \leq
P_e^{u}\big(\{\Sigma_k^r,\Sigma_k^s\}\big) \nonumber \\
& = & Q(\frac{\mu_1 -
\mathcal T_0^*}{\sigma_1}) + \beta Q(\frac{\mathcal T_0^* -
\mu_0}{\sigma_0}) \big|_{\mathbf a = \alpha \cdot \mathbf 1}
\label{eqn:err_3}
\end{eqnarray}
The following theorem summarizes the upper bound of the false alarm
and mis-detection performance.
\begin{Thm} The asymptotic expressions for the false
alarm and mis-detection performance of the proposed cooperative
sensing framework can be upper bounded by
\begin{eqnarray}
\bar{P}_e^{*} & \leq & \frac{1+\beta}{2}\exp \Big(- \frac{K}{3 (1 +
\frac{\Sigma_r}{\rho_r^2})} \nonumber \\
& & \times \big(\frac{1}{\frac{1}{\rho_s} + \sqrt{1
+ \frac{\Sigma_s}{\rho_s^2} + \frac{2}{3\rho_s} +
\frac{1}{\rho_s^2}}}\big)^2\Big) \label{eqn:err}
\end{eqnarray}
where the average SNR of the reporting links $\rho_r = \sqrt{\alpha}
\Lambda_r$ and the average SNR of the sensing links $\rho_s = P_{pu}
\Lambda_s$. \label{Thm:closed_form}
\end{Thm}
\proof Please refer to Appendix \ref{pf:thm_closed_form} for the proof.
\endproof

Theorem \ref{Thm:closed_form} establishes the relation between the
upper bound of the system performance (the false alarm and
mis-detection performance) and the system parameters (the number of
sensor nodes $K$, the channel quality of the reporting links
$\rho_b,\Sigma_b^2$ and the channel quality of the sensing links
$\rho_i,\Sigma_i^2$). In the following subsections, we shall discuss
the relations in details, especially in the low SNR regime.

\subsection{The Effect of $K$} From the expression of
\eqref{eqn:err}, we find that the false alarm and mis-detection performance of the proposed
cooperative sensing framework w.r.t. the number of sensor nodes $K$ scales in the order of
$\mathcal O \big(\exp(-K)\big)$. Hence, to reduce the error performance of the proposed cooperative
scheme by $N$ times, we can simply increase the number of sensor nodes by $\ln N$ times.

\subsection{The Effect of the Reporting Links} To characterize the
effect of the reporting links, we fix the parameter of $\rho_i$ and
$\Sigma_i^2$ and the false alarm and mis-detection error performance
to be $\epsilon$. In the low SNR regime of the reporting links,
equivalently speaking, $\rho_r$ is sufficiently small, the error
expression \eqref{eqn:err} can be approximated as
\begin{eqnarray*}
\epsilon \leq \frac{1+\beta}{2}\exp \Big(- \frac{K}{3 (1 +
\frac{\Sigma_r}{\rho_r^2})} \big(\frac{1}{\frac{1}{\rho_s} + \sqrt{1
+ \frac{\Sigma_s}{\rho_s^2} + \frac{2}{3\rho_s} +
\frac{1}{\rho_s^2}}}\big)^2\Big) \nonumber \\
\approx \frac{1+\beta}{2}\exp \Big(- \frac{K \rho_r^2}{3
\Sigma_r}\big(\frac{1}{\frac{1}{\rho_s} + \sqrt{1 +
\frac{\Sigma_s}{\rho_s^2} + \frac{2}{3\rho_s} +
\frac{1}{\rho_s^2}}}\big)^2\Big) \qquad \
\end{eqnarray*}
We summarize the effect of the reporting links using the following
corollary.
\begin{Cor}[The Effect of the Reporting Links]
With respect to the reporting links, to achieve a target false alarm
and mis-detection performance $\epsilon$ of the proposed cooperative
sensing scheme, the number of sensor nodes $K$ scales in the order
of $\mathcal O\big(\Sigma_r/\rho_r^2\big)$ in the low SNR regime.
\label{cor:for}
\end{Cor}

\subsection{The Effect of the Sensing Links} We apply the same
approach for the sensing links. In the low SNR region of the sensing
links, i.e. $\rho_s$ is sufficiently small, the error expression
\eqref{eqn:err} can be approximated as
\begin{eqnarray*}
\epsilon \leq \frac{1+\beta}{2}\exp \Big(- \frac{K}{3 (1 +
\frac{\Sigma_r}{\rho_r^2})}\big(\frac{1}{\frac{1}{\rho_s} + \sqrt{1
+ \frac{\Sigma_s}{\rho_s^2} + \frac{2}{3\rho_s} +
\frac{1}{\rho_s^2}}}\big)^2\Big) \nonumber \\
\approx \frac{1+\beta}{2}\exp \Big(- \frac{K}{3 (1 +
\frac{\Sigma_r}{\rho_r^2})}\big(\frac{\rho_s}{1 + \sqrt{1 +
\Sigma_s}}\big)^2\Big) \qquad \qquad \quad \ \
\end{eqnarray*}
We summarize the effect of the sensing links using the following
corollary.
\begin{Cor}[The Effect of the Sensing Links]
With respect to the sensing links, to achieve a target false alarm
and mis-detection performance $\epsilon$ of the proposed cooperative
sensing scheme, the number of sensor nodes $K$ scales in the order
of $\mathcal O\big((1 + \sqrt{1 + \Sigma_s})^2/\rho_s^2\big)$ in the
low SNR regime. \label{cor:sen}
\end{Cor}

\section{Simulation Results}
\label{sect:experiments}

\begin{figure}
\centering
\includegraphics[width = 2.5in]{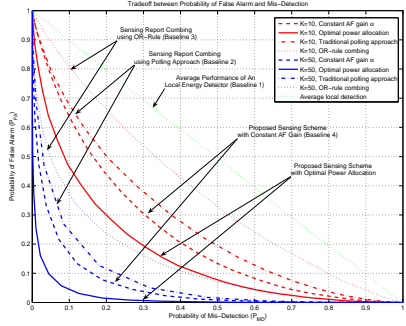}
\caption{Tradeoff curves between the probability of false alarm and mis-detection. The transmit SNR
of the primary user is 30 dB. The mean path loss of the sensing links are $\Lambda_s = -30dB$ with
variance $\Sigma_s = 5dB$. The mean path loss of the reporting links are $\Lambda_r = -80dB$ with
variance $\Sigma_r = 5dB$. As we can see from the above results, the proposed scheme gives a better
tradeoff between the probability of mis-detection and false alarm than other baseline schemes.}
\label{fig:sim_tra}
\end{figure}

\begin{figure}
\centering
\includegraphics[width = 2.5in]{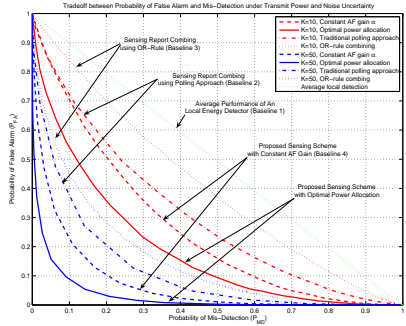}
\caption{Tradeoff curves between the probability of false alarm and mis-detection under transmit
power uncertainties. The mean transmit SNR of the primary user is 30 dB, which is known to all the
sensing nodes. The actual transmit SNR of the primary user is frustrated from 30dB to 50dB and the
exact value is unknown to all the sensing nodes. The mean path loss of the sensing links are
$\Lambda_s = -30dB$ with variance $\Sigma_s = 5dB$. The mean path loss of the reporting links are
$\Lambda_r = -80dB$ with variance $\Sigma_r = 5dB$. The noise variances are frustrated from $-5 dB$
to $5 dB$. As we can see from the above results, under the transmit power uncertainties of the
primary systems, the proposed scheme can still work properly and perform better than other
baselines.} \label{fig:sim_tra_1}
\end{figure}

In this section, we verify our analytical results via simulations. We
assume both the noise variance and the channel statistics are assumed to be constant and known at
the BS for each static period. Each sensor node has the same channel statistics and experience
independent fading. Moreover, each observation chance is chosen to be 2ms and the static period is
assumed to be sufficiently long for 10 observation chances. For easy illustration, we define some
baseline systems, namely the \emph{baseline 1: local detection scheme}, the \emph{baseline 2:
traditional distributed sensing scheme}, the \emph{baseline 3: OR-Rule combining sensing scheme},
and the \emph{baseline 4: proposed cooperative sensing scheme with constant AF gain $\alpha$}. In
baseline 1, we plot the average local detection performance, i.e. the average performance of an
energy detector over different sensing positions. In baseline 2, the BS applies polling scheme to
each of the sensing node for the sensing report and perform the final decision based on majority
voting scheme. In baseline 3, all the sensing node shall report one-bit local decision
simultaneously and the BS performs the final decision based on the OR-Rule. For simplicity, we
assume perfect reporting links for baselines 2 and 3, i.e. the local decision results of baselines
2 and 3 can be successfully delivered to the BS. On the other hand, for our proposed scheme and
baseline 4, the reporting links are modeled by equation \eqref{eqn:sys_mod}. Hence, we have more
favorable assumptions regarding the reporting links for the baselines.  Fig. \ref{fig:sim_tra}
shows the tradeoff relation between the probability of false alarm\footnote{In \cite{Zeng09}, the probability of false alarm $P_{FA}'$ is defined as $\frac{Pr(\hat{\theta} = 1 | \theta = 0) \times Pr (\theta = 0)}{Pr(\hat{\theta} = 1 )}$ and the probability of detection $P_D'$ is defined as $\frac{Pr(\hat{\theta} = 1 | \theta = 1) \times Pr (\theta = 1)}{Pr(\hat{\theta} = 1 )}$. Under these definitions, we can show through our numerical results that $P_{FA}', P_D' \approx 50\%$, which matches the conclusion of \cite{Zeng09}.} $P_{FA}$ and the probability of
mis-detection $P_{MD}$. The proposed cooperative sensing scheme performs better than the local
detection scheme (baseline 1), the traditional distributed sensing scheme (baseline 2) and the
OR-Rule combining sensing schemes (baseline 3) regardless of the power allocation
scheme\footnote{The performance advantage in the tradeoff relations can be translated into quite
period reduction as well. For example, compared with the traditional distributed sensing scheme,
15\% quiet period reduction can be obtained for $P_{MD} = P_{FA} = 0.1$ with 10 sensing nodes.}.
Meanwhile, the proposed cooperative sensing scheme with optimal power allocation achieves better
tradeoff than with constant AF gain scheme (baseline 4). In practise, the actual transmit power of
the primary system may difficult to obtain. In Fig. \ref{fig:sim_tra_1}, we studies the tradeoff
relations when the actual transmit power and the noise variances are unknown to the sensing
nodes\footnote{The actual transmit power frustrated from -10dB to 10dB w.r.t. the mean transmit
power and the noise variances are from -5dB to 5dB as well.}. Based on the numerical results, we
find that the proposed cooperative sensing scheme can still work properly and perform better than
other baselines.

\begin{figure}
\centering
\includegraphics[width = 2.5in]{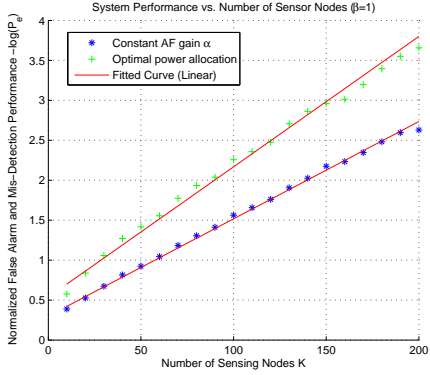}
\caption{System performance versus number of sensor nodes ($\beta =
1$). The simulation points fit very well with the order of growth
expression derived. This verifies that the false alarm and
mis-detection performance decreases exponentially with respect to
the number of sensor nodes $K$ as derived in theorem
\ref{Thm:closed_form}. } \label{fig:sim_nu}
\end{figure}

\begin{figure}
\centering
\includegraphics[width = 2.5in]{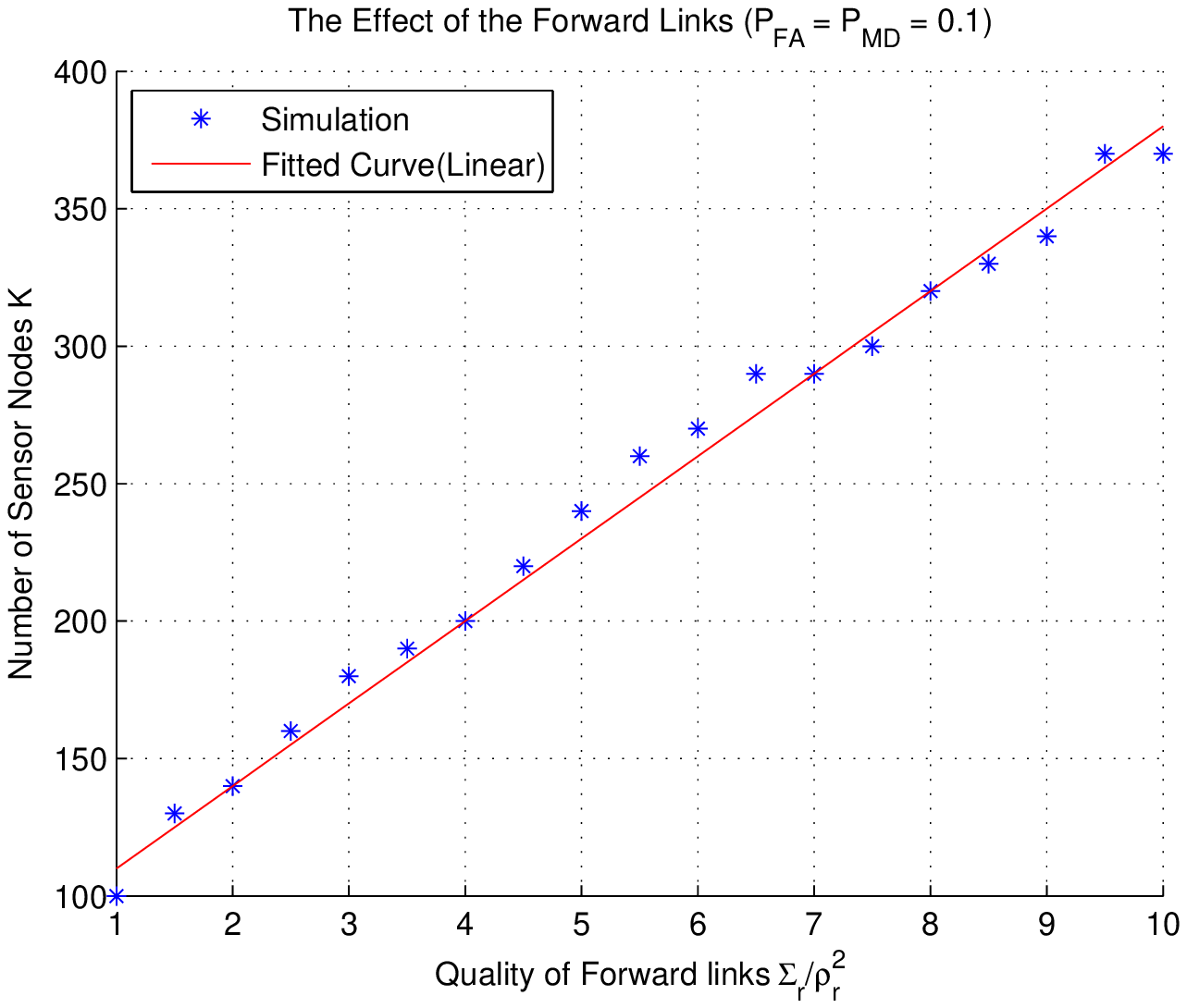}
\caption{The effect of the reporting links ($P_{FA} = P_{MD} =
0.1$). This illustrates the number of sensor nodes required follows
$\mathcal O(\Sigma_r/\rho_r^2)$ as derived in corollary
\ref{cor:for}.} \label{fig:sim_for}
\end{figure}

\begin{figure}
\centering
\includegraphics[width = 2.5in]{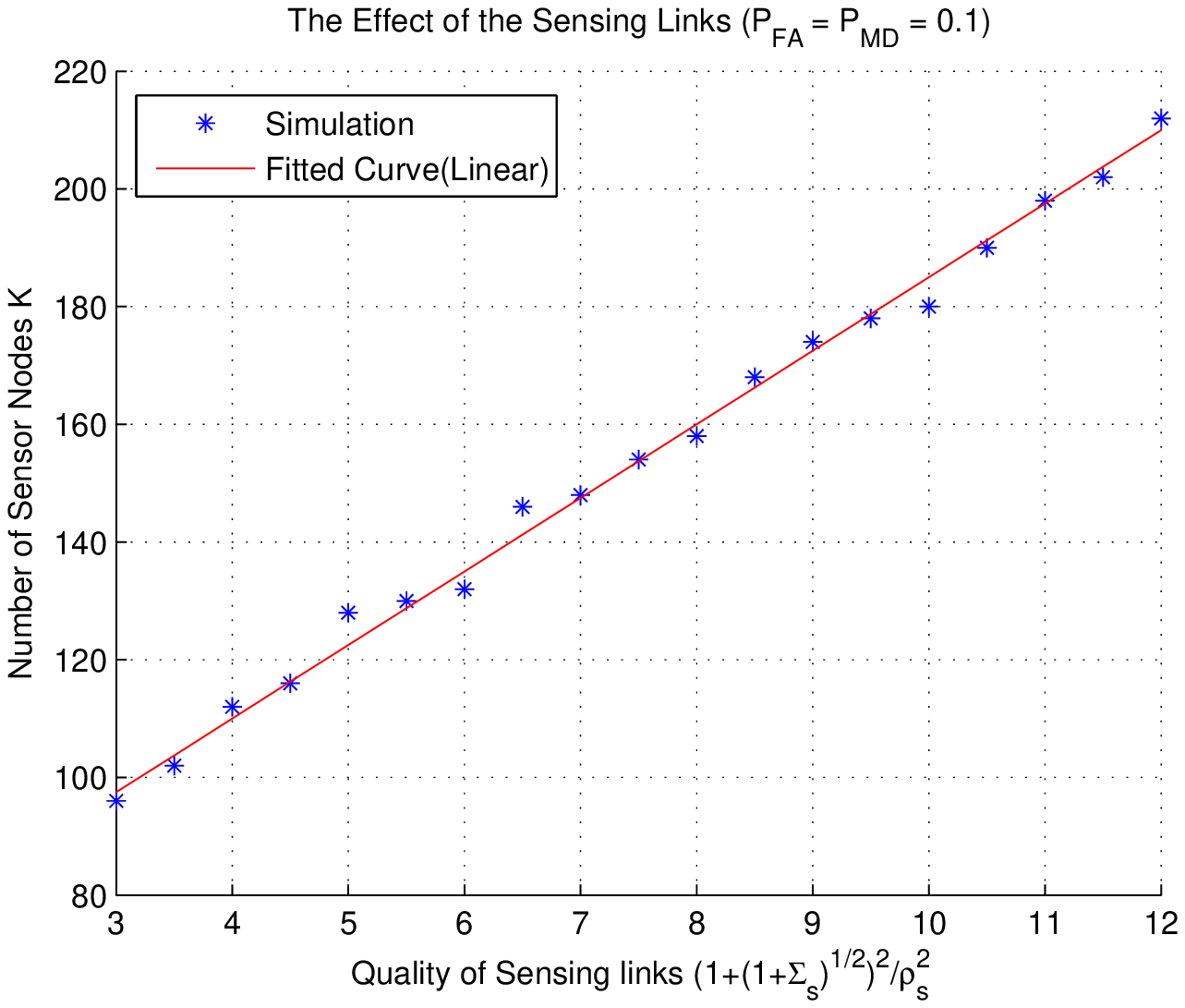}
\caption{The effect of the sensing links ($P_{FA} = P_{MD} = 0.1$).
This illustrates the number of sensor nodes required follows
$\mathcal O\big((1 + \sqrt{1 + \Sigma_s})^2/\rho_s^2\big)$ as
derived in corollary \ref{cor:sen}.} \label{fig:sim_sen}
\end{figure}

Fig. \ref{fig:sim_nu} shows the relations between the system performance (i.e. the false alarm and
mis-detection error performance) and the number of sensor nodes. Without loss of generality, we
choose $\beta$ equals to $1$ and the system performance becomes the sum of the false alarm and
mis-detection error probability. As shown in Fig. \ref{fig:sim_nu}, the system performance of the
proposed cooperative sensing scheme with optimal power allocation as well as the constant AF gain
allocation scales in the order of $\mathcal O\big(\exp(-K)\big)$ as we have shown in Theorem
\ref{Thm:closed_form}. Fig. \ref{fig:sim_for} and Fig. \ref{fig:sim_sen} demonstrate the effects of
the reporting links and sensing links. We simulate the number of sensor nodes required to achieve a
target probability of false alarm and mis-detection (e.g. we choose $P_{FA} = P_{MD} = 0.1$ as
specified in IEEE 802.22 \cite{WRAN06}) under different qualities of the reporting links and
sensing links. As we have shown in the Fig. \ref{fig:sim_for}, the number of sensor nodes required
is proportional to the qualities of the reporting links $\Sigma_r/\rho_r^2$ as derived in Corollary
\ref{cor:for}. Fig. \ref{fig:sim_sen} illustrates the number of sensor nodes $K$ scales in the
order of $\mathcal O\big((1 + \sqrt{1 + \Sigma_s})^2 /\rho_s^2\big)$ w.r.t. the sensing links as
derived in Corollary \ref{cor:sen}.

\section{Conclusion}
\label{sect:conclusion} In this paper, we proposed a simple cooperative sensing framework for the
cognitive radio systems. By applying the proposed cooperative sensing scheme, we formulate the
power scheduling algorithm as a multi-objective optimization problem. To describe the relations of
the multiple objectives, we characterize the optimal trade-off curve, containing a set of Pareto
optimal values for a multi-criteria problem, and derive the asymptotic closed-form expression for
the false alarm and mis-detection error performance of the proposed cooperative sensing framework.
Simulation results are then evaluated to demonstrate the proposed cooperative sensing framework. We
found that the false alarm and mis-detection error performance of the proposed cooperative sensing
framework scales in the order of $\mathcal O \big(\exp(-K)\big)$. In order to achieve a target
false alarm and mis-detection probability, the number of required sensor nodes scales in the order
of $\mathcal O\big(\Sigma_r(1 + \sqrt{1 + \Sigma_s})^2/\rho_r^2\rho_s^2\big)$ in the low SNR
regime.

\appendices
\section{Proof of Lemma \ref{lem:pe}}
\label{pf:lem_pe} In order to calculate the probability of mis-detection and false alarm, we shall
first try to find the probability density function (p.d.f.) of the received signal $X$ at the BS
side. Since the sensing links remain quasi-static throughout the observation chance and the noise
variance is completely known at each sensor node, the time average received signal power at the
$k^{th}$ sensor node can be expressed as\footnote{The proposed framework can be directly extended
to frequency selective channels with OFDM schemes as well.}
\begin{eqnarray}
|x_k|^2  & = & \frac{1}{T}\int_{t=0}^{T} {\big| h_{k}^{s} \theta x_p \sqrt{P_{pu}} + n_{k,t}^{s}\big|}^2
\textrm{d} t \nonumber \\
& = & \frac{1}{T} \Big( |h_{k}^{s} \theta x_p \sqrt{P_{pu}}|^2 T + |n_{k}^{s}|^2 + X_k \Big)\nonumber \\
& {\buildrel \text{w.p.1} \over \longrightarrow} & |h_{k}^{s}|^2 \theta^2 P_{pu} + \frac{1}{T}|n_{k}^{s}|^2
\end{eqnarray}
where $n_{k}^{s}$ is the additive white Gaussian noise with mean zero and variance
$T$\footnote{Without loss of generality, we assume the noise variance is normalized to unity.}. $X_k = 2 \textrm{Re} \Big\{\frac{1}{T}\int_{t=0}^{T}  h_{k}^{s} \theta x_p \sqrt{P_{pu}} n_{k,t}^{s,*} \textrm{d} t\Big\}$ , which is with zero mean and variance $|h_{k}^{s}|^2 \theta^2 P_{pu} T$. For sufficiently large $T$, the cross-term $\frac{X_k}{T}$ tends to zero with probability 1 (w.p.1). We also assume the noise and primary signals are independent and substitute $|x_p|^{2} = 1$ in the last step.

At the BS side, the received signal contributed by $k^{th}$ sensor
node (denoted by $\hat{\mathcal X_k}$) is given by
\begin{eqnarray}
\hat{\mathcal X_k} & = & h_k^b \mathcal X_k = |h_k^r|^2|x_k|^2\sqrt{\alpha_k} \nonumber \\
& = & \underbrace{|h_k^r|^2
|h_k^s|^2 \theta^2 P_{pu} \sqrt{\alpha_k}}_{\textrm{Signal Part}} + \underbrace{\frac{1}{T}|h_k^r|^2
|n_{k}^{s}|^2 \sqrt{\alpha_k}}_{\textrm{Noise Part}}
\end{eqnarray}
The statistical properties of the random variable $\hat{\mathcal
X_k}$ can be evaluated in the following way.
\begin{eqnarray}
\mu_{\hat{\mathcal X_k}} & = & \mathbb
E_{h_k^r,h_k^s,n_k^s}\big[\hat{\mathcal X_k}\big] =
\Sigma_k^r \sqrt{\alpha_k} \big(\Sigma_k^s \theta^2 P_{pu} + 1 \big)\\
\sigma_{\hat{\mathcal X_k}}^2 & = & \mathbb E_{h_k^r,h_k^s,n_k^2}
\big[(\hat{\mathcal X_k}- \mu_{\hat{\mathcal X_k}})^2\big] \nonumber \\
& = & \Sigma_{k}^{r,2} \alpha_k (3 \Sigma_{k}^{s,2} \theta^4 P_{pu}^2
+ 2\Sigma_{k}^{s} \theta^2 P_{pu}+ 3)
\end{eqnarray}

Using the proposed cooperative sensing scheme, all $K$ sensor nodes
are suggested to report their local measurements at the same time
and all the reports will naturally combined in the air interface.
Applying the central limit theorem \cite{Papoulis02}, the received
signal $X$ at the BS side can be approximated by the following
equation when $K$ is sufficiently large \cite{Tandra08}
\begin{eqnarray}
X \sim \mathcal N \bigg( \frac{\sum_{k=1}^{K} \mu_{\hat{\mathcal
X_k}}}{K}, \frac{\sum_{k=1}^{K}\sigma_{\hat{\mathcal X_k}}^2 +
1}{K^2} \bigg)
\end{eqnarray}
where $\mathcal N (\mu,\sigma^2)$ denotes the Gaussian distribution
with mean $\mu$ and variance $\sigma^2$.

Denote $\mu_1, \sigma_1^2$ to be the mean and variance of the
received signal $X$ when the primary user is in active state and
$\mu_0, \sigma_0^2$ to be the mean and variance of $X$ when the PU
is in idle state, respectively. We can derive the expressions for
$\mu_0,\sigma_0^2,\mu_1,\sigma_1^2$ as follows.
\begin{eqnarray}
\mu_0 & = & \frac{\sum_{k=1}^{K} \mu_{\hat{\mathcal X_k}}}{K} = \frac{\sum_{k=1}^{K} \Sigma_k^r
\sqrt{\alpha_k} }{K}, \\
\sigma_0^2 & = & \frac{\sum_{k=1}^{K} \sigma_{\hat{\mathcal
X_k}}^2}{K^2} = \frac{3 \sum_{k=1}^{K} {\Sigma_k^r}^2 \alpha_k  + 1}{K^2} \\
\mu_1 & = & \frac{\sum_{k=1}^{K} \mu_{\hat{\mathcal X_k}}}{K} =
\frac{\sum_{k=1}^{K}\Sigma_k^r \sqrt{\alpha_k}(\Sigma_k^s P_{pu} + 1)}{K} \\
\sigma_1^2 & = & \frac{\sum_{k=1}^{K} \sigma_{\hat{\mathcal
X_k}}^2}{K^2} \nonumber \\
& = & \frac{ \sum_{k=1}^{K}{\Sigma_k^r}^2 \alpha_k (3
\Sigma_k^{s,2} P_{pu}^2 + 2 \Sigma_k^s P_{pu} + 3) + 1}{K^2}
\end{eqnarray}

Using these approximations gives the following expressions, we have $Pr \big(\hat{\theta} = 0 |
\theta = 1, \{\Sigma_k^r,\Sigma_k^s\}, \mathbf a, \mathcal T_0 \big) = 1 - Q \big(\frac{\mathcal
T_0 - \mu_1}{\sigma_1}\big)$ and $Pr \big(\hat{\theta} = 1 | \theta = 0, \{\Sigma_k^r,\Sigma_k^s\},
\mathbf a, \mathcal T_0 \big) = Q \big( \frac{\mathcal T_0 - \mu_0}{\sigma_0} \big) $, where
$Q(\cdot)$ denotes the standard Gaussian complementary c.d.f.\cite{Proakis00}. Notice that $1 -
Q(-x) = Q(x)$, we have Lemma \ref{lem:pe}.

\section{Proof of Theorem \ref{Thm:pri_dual}}
\label{pf:thm_pri_dual} Since the primal problem \eqref{eqn:thr_4}
is not convex, the standard optimization theory cannot be applied
here to prove the zero duality gap. From the result of \cite{Yu06},
the primal problem and its dual problem will have zero duality gap
when the \emph{time-sharing} condition is satisfied. The results
will hold true even when the primal problem is nonconvex.

Let $P_{1} = \{P_{0,1},P_{1,1},\ldots,P_{K,1}\},P_{2} =
\{P_{0,2},P_{1,2},\ldots,P_{K,2}\}$ and $P_{3} =
\{P_{0,3},P_{1,3},\ldots,P_{K,3}\}$ be values of power constraints
with $P_{3} = \nu P_{1} + (1-\nu) P_{2}$ for some $0 \leq \nu \leq
1$. Let $\mathbf a_1^{*}$, $\mathbf a_2^{*}$ and $\mathbf a_3^{*}$
be the optimal power allocation (AF gain) scheme to the primal
optimization problem \eqref{eqn:thr_2} with constraints
$P_{1},P_{2}$ and $P_{3}$, respectively. Let $P_{e1}^{*}$ and
$P_{e2}^{*}$ be their respective optimal values of the false alarm
and mis-detection error performance. To prove the time-sharing
property, we need to construct a power allocation (AF gain) scheme
$\mathbf a_3^{*}$ such that it achieves an error equal to or lower
than $\nu P_{e1}^{*} + (1- \nu)P_{e2}^{*}$ with a power that is at
most $\nu P_{1} + (1- \nu) P_{2}$ for all $\nu$ between zero and
one. Such a scheme $\mathbf a_3^{*}$ may be constructed as follows.

Without loss of generality, we consider the multiple transmission opportunities are divided into
two periods with $\nu$ proportion of which corresponding to a power constraint $P_{1}$ and
$(1-\nu)$ proportion of which corresponding to a power constraint $P_{2}$. We can apply the power
allocation scheme $\mathbf a_{1}^{*}$ to the first $\nu$ proportion of channel realizations and
$\mathbf a_2^{*}$ to the rest $(1 - \nu)$ proportion. By doing so, the system achieves the value of
$(\mathcal T_0^{*} - \mu_0)/\sigma_0$ at least equal to $\nu P_{e1}^{*} + (1- \nu)P_{e2}^{*}$ and
therefore, the time-sharing property holds. From \cite[Theorem 1]{Yu06}, the primal problem
\eqref{eqn:thr_2} and the dual problem \eqref{eqn:tar_3} have the same optimal value since the
time-sharing condition is satisfied.

\section{Proof of Theorem \ref{Thm:closed_form}}
\label{pf:thm_closed_form} From the expressions of $\mu_0,\mu_1,\sigma_0^2$ and $\sigma_1^2$, we
find that when the number of sensor nodes $K$ is sufficiently large\footnote{In the numerical
examples, we found that the first expression is quite accurate for moderate $K \sim 20$.}, the
following relation holds, $(\mu_1 - \mu_0)^2 \gg 2(\sigma_1^2 - \sigma_0^2) \ln(\beta
\sigma_1/\sigma_0)$. Using the above relation, the optimal threshold $\mathcal T_0^{*}$ from
\eqref{eqn:thre_1} is given by
\begin{eqnarray}
\mathcal T_0^{*} & = & \mu_0 + \frac{\sigma_0^2 (\mu_0 - \mu_1)}{\sigma_1^2 - \sigma_0^2} + \frac{\sigma_0 \sigma_1}{\sigma_1^2 - \sigma_0^2}
\nonumber \\
& &  \times \sqrt{(\mu_1 - \mu_0)^2 + 2(\sigma_1^2 -
\sigma_0^2) \ln(\beta \sigma_1/\sigma_0)} \\
& \approx &\mu_0 + \frac{\sigma_0^2 (\mu_0 - \mu_1) + \sigma_0 \sigma_1 \sqrt{(\mu_1 -
\mu_0)^2}}{\sigma_1^2 - \sigma_0^2} \\
& = &\mu_0 + \sigma_0 \frac{\mu_1 - \mu_0}{\sigma_0 + \sigma_1}
\label{eqn:err_4}
\end{eqnarray}
Substitute the equation \eqref{eqn:err_4} into \eqref{eqn:err_3}, we
have
\begin{eqnarray}
P_e^{u}\big(\{\Sigma_k^r,\Sigma_k^s\}\big) = (1 + \beta) Q(\frac{\mu_1 - \mu_0}{\sigma_0 +
\sigma_1}) \big|_{\mathbf a = \alpha \cdot \mathbf 1} \label{eqn:err_5}
\end{eqnarray}
and the average probability of false alarm and mis-detection is
\begin{eqnarray}
\bar{P}_e^* & \leq & \mathbb E_{\{\Sigma_k^r,\Sigma_k^s\}} \Big[
P_e^{u}\big(\{\Sigma_k^r,\Sigma_k^s\}\big) \Big] \nonumber \\
& = & \mathbb E_{\{\Sigma_k^r,\Sigma_k^s\}}\Big[ (1 +
\beta) Q(\frac{\mu_1 - \mu_0}{\sigma_0 + \sigma_1})\big|_{\mathbf a = \alpha \cdot \mathbf 1} \Big]
\label{eqn:ana_1}\\
& \leq & \mathbb E_{\{\Sigma_k^r,\Sigma_k^s\}}\bigg[ \frac{1+\beta}{2} \exp \Big( -
\frac{1}{2}\big(\frac{\mu_1 - \mu_0}{\sigma_0 + \sigma_1} \big)^2 \Big) \big|_{\mathbf a = \alpha
\cdot \mathbf 1} \bigg] \nonumber \\
& \leq & \frac{1+\beta}{2} \exp \big( - \frac{\gamma^2}{2} \big)
\label{eqn:err_6}
\end{eqnarray}
with $\gamma =  \mathbb E_{\{\Sigma_k^r,\Sigma_k^s\}} \bigg[ \frac{\mu_1 - \mu_0}{\sigma_0 +
\sigma_1} \big|_{\mathbf a = \alpha \cdot \mathbf 1} \bigg] $ where we use the relation $Q(x) \leq
\frac{1}{2}\exp(-\frac{x^2}{2})$ in the second inequality and the concavity of the function
$\exp(-\frac{x^2}{2})$ w.r.t. $x$ in the last step.

Applying the statistical model given by \eqref{eqn:mod_1} to
\eqref{eqn:mod_4}, the mean and variance of the received
$\textrm{SNR}_{k}^{r}$ and $\textrm{SNR}_{k}^{s}$ under the constant
AF gain scheme are given by $(\sqrt{\alpha}\Lambda_r, \Sigma_r)$ and
$(P_{pu} \Lambda_s, \Sigma_s)$, respectively.
$\mu_0,\mu_1,\sigma_0^2$ and $\sigma_1^2$ under the constant AF gain
policy $\mathbf a = \alpha \cdot \mathbf 1$ are thus given by
\begin{eqnarray*}
\mu_0 & = & T \sqrt{\alpha} \Lambda_{r}, \\
\sigma_0^2 & = & \frac{3 T^2 (\alpha \Lambda_r^2 + \Sigma_r)}{K}, \\
\mu_1 & = & T \sqrt{\alpha} \Lambda_{r} (P_{pu} \Lambda_s+ 1), \\
\sigma_1^2 & = & \frac{ T^2 (\alpha \Lambda_r^2 + \Sigma_r) \big(3
P_{pu}^2 \Lambda_s^2 + 3 \Sigma_s + 2 P_{pu} \Lambda_s + 3 \big)}{K}
\end{eqnarray*}
where we have substituted the mean and variance of the received
$\textrm{SNR}_{k}^{b}$ and $\textrm{SNR}_{k}^{i}$. Hence, we can
evaluate the expression of $\gamma$ as follows.
\begin{eqnarray}
\gamma \approx \sqrt{\frac{K}{3(1 + \frac{\Sigma_r}{\alpha
\Lambda_r^2})}}\cdot\frac{P_{pu} \Lambda_s}{1 + \sqrt{ P_{pu}^2
\Lambda_s^2 + \Sigma_s + \frac{2P_{pu} \Lambda_s}{3} + 1}}
\label{eqn:gamma}
\end{eqnarray}

Combine \eqref{eqn:err_6} and \eqref{eqn:gamma}, we have Theorem \ref{Thm:closed_form}.

\bibliographystyle{IEEEtran}
\bibliography{IEEEabrv,dis_sen_v2.6}

\end{document}